\documentstyle[trans]{rspub1}
\input epsf.sty
\begin{document}

\title{Thin-Film Trilayer Manganate Junctions}
\author{J. Z. Sun}
\affiliation{IBM T. J. Watson Research Center\\P. O. Box 218\\Yorktown
Heights, NY 10598, U. S. A.}
\maketitle

\begin{abstract}
Spin-dependent conductance across a manganate-barrier-manganate junction has
recently been demonstrated. The junction is a La$_{0.67}$Sr$_{0.33}$MnO$_3$%
-SrTiO$_3$-La$_{0.67}$ Sr$_{0.33}$MnO$_3$ trilayer device supporting
current-perpendicular transport. Large magnetoresistance of up to a factor
of five change was observed in these junctions at 4.2K in a relatively low
field of the order of 100 Oe. Temperature and bias dependent studies
revealed a complex junction interface structure whose materials physics has
yet to be understood.
\end{abstract}

\label{firstpage}

\section{Introduction}

Since the discovery of large magnetoresistance (MR) at room temperature in La%
$_{0.67}$Ba$_{0.33}$MnO$_3$ epitaxial thin films \cite{71}, there has been a
resurgence of interest in doped perovskite manganate compounds. Doped
perovskite manganates in this case refers to the family of materials related
to compounds with a chemical composition of A$_{1-x}$B$_x$MnO$_3$, where A
stands for trivalent rare-earth such as La, Pr, etc. and B stands for
divalent alkaline-earth elements, such as Ba, Sr, etc. Materials refinements
have revealed that very large magnetoresistances, of over three orders of
magnitude change in resistance, can be observed in such compounds prepared
through special heat-treatment \cite{66,64,62}. These materials have since
been refered to as colossal magnetoresistance materials, or CMR materials.
The large magnetoresistance effect in these compounds was referred to as
CMR\ effect.

The physics of CMR\ is complex due to the similar
energy scales involved for interactions among the lattice, electronic and
magnetic degrees of freedom. The basic electronic interactions in these
materials were investigated back in the fifties. The double-exchange model 
\cite{81,82} was proposed to explain the simultaneous onset of metallic
conductivity and ferromagnetism as doping concentration $x$ is increased across
the insulator-metal transition point of $x\sim 0.2$. It was also understood
then that there is a strong correlation between the Mn-O-Mn bond length
and bond angle and the magnetic coupling of adjacent Mn ions \cite{91}. A
Jahn-Teller distortion lifts the double-degeneracy of Mn $d$-electrons' $e_g$
orbitals, and provides a mechanism for strong coupling between the
electronic, magnetic and lattice degrees of freedom. In bulk ceramic
materials a Jahn-Teller distortion-related orthorombicity is present in
La$_{1-x}$Ca$_x$MnO$_3$ in the doping range of $0\leq x \leq 0.2$.\cite{106}.
Above $x=0.2$ such static distortion of lattice is no longer observable.

To explain the CMR effect, recently it was proposed that even for $x>0.2$, 
dynamically fluctuating local Jahn-Teller distortion still
exists, which provides a mechanism for the localization of ferromagnetic
polarons \cite{41,49,176}. For CMR\ materials at $x\sim 0.3$ in
temperatures above the Curie point $T_c$, these polarons are localized.
Electrical conduction occurs via hopping process. For temperatures well
below $T_c$ on the other hand, the electornic states appear more extended,
and a band-like conduction is more appropriate. Local Spin-Density
Approximation calculation \cite{19} has been carried out for La$_{1-x}$Ca$_x$%
MnO$_3$ in the region $x=\frac 14\sim \frac 13$. Their results suggest an
almost completely spin-polarized Mn $d$-band. Therefore these materials may
behave as half-metals at temperatures below $T_c$, and hence may exhibit a
large spin-dependent conductance across a tunneling barrier \cite{28}.

For magnetic field-sensing applications, low-field responsivity is
necessary. Large MR is desired in field ranges of tens to hundreds of Oe.
This presents a challenge for CMR materials, since CMR effect observed in
the generic doped manganate perovskites involves a magnetic field above 1
tesla \cite{66,71}, and the low-field magnetoresistance in the field range
of 10-100Oe remains minimal.

One might expect large spin-dependent conductivity across a
macroscopic interface between two CMR electrodes across which the
magnetization abruptly changes direction. This may either be due to
local spin-dependent hopping, as prescribed by the double-exchange
mechanism, or it maybe due to spin-dependent tunneling, if the
interface is electrically insulating enough to behave as a tunneling
barrier. Such a macroscopic interface disrupts the magnetic
exchange-coupling, so that easy rotation of magnetic moment from one
electrode to the other can be obtained.

Three approaches have been taken to experimentally exploit this concept. The
first approach is to use a crystalline grain-boundary as the interface, and
study the spin-dependent transport across. The role of grain boundary in
providing additional contributions to low-field magnetoresistance has long
been suspected \cite{254,174,198}. Direct experimental observation of
grain-boundary-originated magnetoresistance has recently been made \cite
{344,366} which shows a 30\% magnetoresistance in a field of less than 500
Oe.

The second approach uses the naturally occuring inter-planar conductivity in
the 2-dimensional version of the CMR\ material La$_{1.4}$Sr$_{1.6}$Mn$_2$O$%
_7 $ \cite{242}. They demonstarted a large magnetoresistance of
240\% in less than 500Oe at 4.2K, using the layered single crystal
compound of La$_{2-2x}$Sr$_{1+2x}$Mn$_2$O$_7$ at $x=0.3$. The transport
direction was perpendicular to the planes of layering, and the authors
argued that transport in this direction may be due to spin-dependent
tunneling.

The third approach makes use of thin film trilayer junctions, fabricated
from epitaxial trilayer films to support current-perpendicular (CPP) 
transport. Successful
demonstration of large low-field magnetoresistance in CMR-based devices was
first realized in such junctions \cite{142,191}. The
devices are lithographically fabricated from La$_{0.67}$(Ca,Sr)$%
_{0.33}$MnO$_3$ - SrTiO$_3$ - La$_{0.67}$(Ca,Sr)$_{0.33}$MnO$_3$ (LSMO - STO
- LSMO, or LCMO - STO - LCMO) trilayer epitaxial thin films. The junctions
show large low field magnetoresistance, of about a factor of two to five, in
a field around 100Oe at 4.2K. These results provide an existence proof that
low-field CMR is possible.

This paper focuses on experimental issues concerning spin-dependent
transport in CMR trilayer junctions.

\section{Structure of a thin film trilayer junction}

\subsection{Conceptual design of the device}

\begin{figure}[tbp]
\epsfxsize=6in
\epsfbox{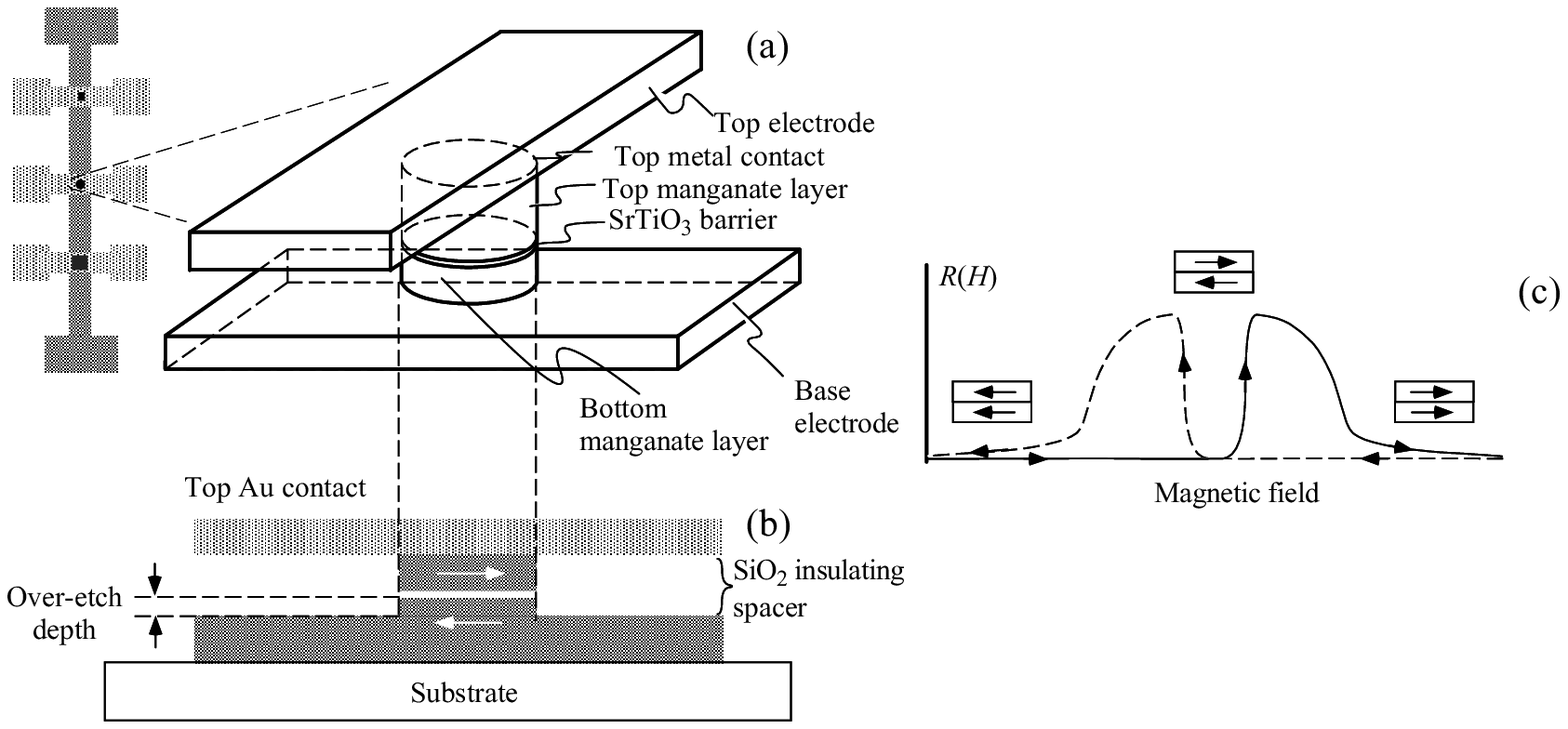}
\caption{A\ schematic view of the LSMO-barrier-LSMO\ trilayer thin film
junction structure. (a) Left:\ top-view of the device; right: 3-dimensional
illustration of the current-perpendicular pillar structure. (b): Side-view
of the structure, showing the over-etch step which adds additional magnetic
coupling between the top and bottom ferromagnetic electrodes. (c) Junction
resistance as a function of sweeping magnetic field, showing the transitions
from parallel to anti-parallel to parallel state of the magnetic moment
alignments of the electrodes. Figure reproduced from ref.\protect\cite{364}}
\label{fig1}
\end{figure}

A typical structure for thin film trilayer magnetic junction is illustrated
in Fig.\ref{fig1}. The junction is comprised of a top and a bottom magnetic
electrode, made of epitaxial CMR\ thin films. The two electrodes are
separated in between by a thin layer of foreign material such as SrTiO$_3$,
which disrupts the magnetic exchange coupling between the electrodes, and
makes it possible to macroscopically rotate the magnetic moment of one
electrode with respect to another. Transport current is forced to flow
perpendicular to barrier in the CPP geometry. If the barrier is thin enough,
it will allow some passage of electrical current, either via
metal-insulator-metal tunneling, or by some other more complex and
inhomogeneous processes, such as defect-assisted hopping through the
barrier. One hopes to directly observe spin-dependent transport across the
barrier. By spin-dependent transport we mean that, on a macroscopic scale,
the conductance across the barrier is dependent on the relative orientation
of the ferromagnetic (FM) moments of the two electrodes. Usually, the
transport resistance across such a structure would be minimum when the two
FM electrodes' moments are parallel, and maximum when they are
antiparallel. This is true in many mechanisms including spin-dependent
tunneling, double-exchange-mediated nearest neighbor hopping, or in metallic
two-channel spin-scattering limited conduction.

When such a trilayer is subjected to an applied magnetic field, the junction
resistance will show a particular type of field dependence. The junction
stays in its low resistance state in sufficiently high field. But when the
applied field is swept from its negative value to positive, one electrode
will magnetically rotate before the other if the device is designed such
that the two electrodes have different magnetic anisotropy. This will result
in a momentary antiparallel arrangement of the relative moments across the
barrier, and the junction resistance will register its high value in this
field range. This is schematically illustrated in Fig.\ref{fig1}(c).

The conduction mechanism across the barrier layer can vary. In metal or
alloy-based thin film trilayers, several types of separation layers have
been experimented with. The separation layer can either be a non-magnetic
metal, as in the case of a spin-valve \cite{372}, or an insulator, as in
spin-dependent tunneling \cite{28,362,mr6,363}. Spin-dependent tunneling has
been observed in elemental and alloy ferromagnetic metal electrode systems 
\cite{362,206,223,25,27,134,222,238}. There are still debates as to whether
such type of tunneling should be observable in $d$-band dominant CMR
materials \cite{348}, since $d$-electrons decay much more rapidly into the
barrier than $s$-electrons \cite{346,347}.

For CMR\ trilayers, the most commonly used barrier material is a thin layer
of epitaxial SrTiO$_3$ film, 20 to 50\AA\ thick. The role of this barrier
material is yet to be fully understood, as will be further discussed later.

Non-linear current-voltage characteristics have been observed in
CMR-barrier-CMR junctions \cite{142,293,191}. These current-voltage ($IV$)
characteristics appear roughly consistent with a metal-barrier-metal
tunneling process. However, non-linear $IV$ characteristic alone is
insufficient for establishing a tunneling transport mechanism, and other
mechanisms have been known to give rise to non-linear $IV$s as well \cite
{335,353,354}. The exact conduction mechanism in these CMR\ junctions
remains unclear at present.

\subsection{Fabrication}

In our lab, thin film trilayers were made using epitaxial La$_{0.67}$Sr$%
_{0.33}$MnO$_3$ (LSMO) electrodes and SrTiO$_3$ (STO) barriers \cite
{191,142,293}. Briefly, epitaxial growth was achieved by using {\it in situ}
pulsed-laser deposition. Typical growth conditions included a substrate
temperature of 600 to 800C, and an oxygen pressure of 300mTorr. A\ Nd-YAG
laser was used in its frequency tripled mode (355nm) at 10Hz repetition. The
laser intensity on target surface was estimated to be around 3 to 5 Joules/cm$%
^2$. Deposition was done using the sequence of 600\AA\ LSMO, followed by 20
to 50\AA\ of STO, followed again by 400\AA\ of LSMO. Typical deposition rate
for LSMO\ was around 2.6\AA /sec, STO about 1.8\AA /sec. Single crystal
substrates of SrTiO$_3$(100), LaAlO$_3$ (100)\ and NdGaO$_3$ (110) have all
been used for experimentation. After deposition, the substrate was cooled in
300Torr of oxygen to room temperature in about 1 hr. A\ thin layer of
silver, typically 500\AA\ or less, was sputter deposited to the surface of
the trilayer before it was taken out of the vacuum system for
further processing.

Structurally, on the local scale of several thousand angstroms at least, the
LSMO/STO/LSMO\ interfaces are well-formed, and are free of gross defects.
The continuation of epitaxial growth of LSMO\ across the STO\ barrier layer
is confirmed by X-ray diffraction as well as by cross-sectional transmission
electron microscopy (TEM). A representative TEM picture of the trilayer
interface is shown in Fig.\ref{fig2}, in which case a coherent lattice
fringe can be seen to carry across the STO\ barrier layer from the bottom
LSMO\ to the top.

Optical photolithography was used to fabricate CPP\ junctions from trilayers.
The shape of the base-electrode layer was formed first, by ion
milling through a photoresist stencil. Ion milling was done using 
neutralized Ar ions, 500eV, 0.3mA/cm$^2$, 45$^o$ incidence angle. After
stripping the resist, a second resist pattern was put on, which defined the
pillar structure forming the junctions. Again ion milling was used, in this
case to etch half way into the film, timed to stop immediately past the STO\
barrier layer, forming the pillar structure from the trilayers. A blanket SiO%
$_2$ layer, typically about 3000\AA\ thick, was then sputter deposited on
top, followed by a lift-off step that removes the resist layer on top of the
pillars, opening up self-aligned holes for top metallic contact. A\ layer of
about 2000\AA\ sputtered gold was used for top contact, which was
subsequently etched using yet another level of photoresist step, forming the
cross-bar structure from which 4-probe measurements could be made of
individual junction pillars underneath each gold contact bridge. A scanning
electron micrograph (SEM) of a typical device structure made with this
process is shown in Fig.\ref{fig4}. To make certain there is no
interdiffusion-related insulation problem between SiO$_2$ and gold, as well
as for promoting adhesion, a thin layer of titanium, usually about 20 to 50
\AA\ thick, was put down between the SiO$_2$ and the top gold layer. The
quality of SiO$_2$ insulation was further confirmed by the consistency
between results from devices fabricated using MgO insulation layer and from
devices made using SiO$_2$.

\section{Transport Properties}

\subsection{Overview}

Two types of junction behaviors were observed. They can be 
classified according to their temperature dependence. Fig.\ref{fig5}
presents a summary. In one type of junctions, typically with STO\ barriers
on the thicker side (30 to 50\AA ), the junction resistance increases with
decreasing temperature, as shown in Fig.\ref{fig5}(a). These typically
involve junctions of sizes below 10$\mu m$, and are devices with the largest
observed magnetoresistance at low temperatures. For this type of junctions,
the magnetic field dependence of junction resistance, $R(H)$, tends to be
rather complex, an example is shown in Fig.\ref{fig5}(b).

Another type of junctions, typically involving thinner STO\ barriers (below
30\AA ), shows a decreasing junction resistance upon the lowering of sample
temperature. These junctions often have cleaner, more reproducible $R(H)$
loops, as shown in Fig.\ref{fig5}(d). The interpretation of data from these
low-resistance junctions is tricky, because the junction resistance could 
approach the base-electrode's sheet resistance. For a 600\AA\
LSMO\ base electrode with resistivity of $10^{-4}$ to $10^{-3}\Omega cm$ at
4.2K, we have a base-electrode $R_{\Box }\approx 15$ to $150\Omega $. When
junction resistance approaches that of $R_{\Box }$, distributed voltage drop
in the base electrode can no longer be ignored, and one has to be
careful in distinguishing between true junction resistance and the apparent
resistance caused by voltage distribution inside the base electrode \cite
{383}. Such artifacts can cause a superficially high value of
magnetoresistance \cite{216}.

Another difficulty in studying low resistance junctions has to do with
sample heating. It is difficult to apply a large voltage across a low
resistance junction without significantly heating up the {\it junction area}
of the sample. Perovskites such as SrTiO$_3$ have relatively low thermal
conductivity (around 10 W/Km for STO at room temperature, compared to about
144 W/Km for single crystal silicon, for example. NdGaO$_3$ single crystal
substrates have probably even lower thermal conductivity). Therefore, sample
heating can be significant on a local scale for the junction, even when the
input power is insufficient to cause much observable heating over the chip).
For these two reasons, most of our discussions will be focused on devices of
the first type, namely the high resistance junctions whose resistance
increase upon cooling of the sample, and whose resistance value is at least
an order of magnitude larger than the base-electrode $R_{\Box }$.

\subsection{Temperature dependence of junction resistance}

A\ representative temperature dependence of junction resistance is shown in
Fig.\ref{fig6}. Two regions of temperatures involving distinctively
different junction behaviors can be identified. In the first region,
typically between 130K and 300K, the junction resistance rises almost
exponentially upon the decreasing of temperature, resembling a $T^{1/4}$
scaling relation commonly seen in systems exhibiting variable range hopping 
\cite{237}. A replot of the junction resistance vs. temperature data
according to the scaling relation $R\left( T\right) =R_o\exp \left[ \left(
T^{*}/T\right) ^{1/4}\right] $ is shown in the inset of Fig.\ref{fig6}. In
this region the magnetoresistance is minimal, typically less than 0.5\% for
a field sweep of couple of hundred Oe.

Below 130K, the junction enters its second region, where the junction
resistance is less sensitive to temperature variation. It becomes more
noisy, and exhibits large magnetic field-exposure history dependence. This
is the temperature region where large low-field magnetoresistance is
observed. An example is shown in Fig.\ref{fig7}, where the junction
resistance is measured as a function of sweeping magnetic field at a
constant temperature of 4.2K. Such response to sweeping magnetic field is
expected conceptually from the simple two-FM-electrode junction picture as
shown in Fig.\ref{fig1}(c). Details about the shape of the $R(H)$ loops will
be discussed in the next section, here we merely use Fig.\ref{fig7} to point
out that the low-field magnetoresistance is large. For the junction shown
here, a factor of 5 change in resistance is observed at a switching field of
around 100 Oe at 4.2K. This is the first observation of large MR\ at
low-fields in CMR\ materials, proving that CMR\ is not a phenomenon limited
to high magnetic fields.

We define two resistance values from data such as shown in
Fig.\ref{fig7}, namely $R_{high}$ and $R_{low}$, corresponding to the
dc resistive high- and low-state of the junction. The
magnetoresistance contrast can then be conveniently defined as
$C_R=(R_{high}-R_{low})/R_{low}$, according to the conventions used in
literatures. The temperature dependences of $R_{high}$, $R_{low}$ and
$C_R$ are shown in Fig.\ref{fig5}, Fig.\ref{fig6} and Fig.\ref{fig8}. As mentioned
above, the magnetoresistance decreases with increasing temperature,
and it disappears around 130K. This is well below the Curie
temperature $T_c$ of the thin film material which is close to the
resistance peak temperature. The resistance peak temperature of the
base electrode is shown in Fig.\ref{fig8} to be above 300K. Separate
measurements confirm the base-electrode's resistance peaks around
350-360K. Clearly, something other than the film's $T_c$ is limiting
the upper temperature of the junction magnetoresistance.

The high temperature behavior of junction resistance as shown in
Fig.\ref {fig6} may suggest defect-assisted hopping as a possible
conduction mechanism across the STO\ barrier. Indeed such temperature
dependence of resistance is common in oxide perovskite semiconductors
having a large population of defect sites. The defect assisted
conduction process may provide additional high temperature conduction
channels that progressively shunt out the direct, spin-dependent
conduction process manifested at low temperatures. For this scenario
to work, the defect-assisted hopping has to have a large spin-flipping
rate, because the high temperature conduction apparently has lost all
spin-dependence information, judging from its minimal value of
magnetoresistance.

There are other possibilities for a premature decrease of
magnetoresistance. Spin-dependent transport across a trilayer is
sensitive to interface magnetic states which may be different from
that deep inside the film. This may be caused by a reduced magnetic
coupling at the interface, by a discontinuous electronic structure, or
by interface diffusion or oxygen deficiency. A different stress
environment at the interface may cause change in electrode's magnetic
states. Excitation of surface magnons have also been suggested as a
possible source for the suppression of magnetoresistance, both for
high temperature and for elevated junction bias\cite{388}.

\subsection{Magnetic field dependence of junction resistance}

The magnetic coercivity of 33\%-doped LSMO\ thin films is of the order of
100 to 200 Oe at helium temperature \cite{345,364}. In CMR\ junctions, the
situation is more complex. Additional magnetic interactions need to be
considered. Four major contributors are: the shape anisotropy of the
junction pillar, the edge-coupling, the dipolar coupling across a rough
junction interface, and the magnetostriction-induced anisotropy. The
complexity of magnetic states in the junction pillar is apparent from our
measurement such as data shown in Fig.\ref{fig7}.

Junctions with high resistance have more complex $R(H)$ loops. In Fig.\ref
{fig9}, $R(H)$ loops from two junctions of different resistance are compared
to the magnetic hysteresis loop of a blank film of La$_{0.67}$Sr$_{0.33}$MnO$%
_3$. The top panel shows a representative magnetic hysteresis loop taken on
a blank La$_{0.67}$Sr$_{0.33}$MnO$_3$ epitaxial film. The middle panel (b)
shows the $R(H)$ loop for a low-resistasnce junction. The switching field
for this type of junction agrees well with that of the coercivity of the
blank film. The lower panel (c) shows the $R(H)$ loop of a high-resistance
junction. Here the lower switching field agrees well with the magnetic
coercivity of the blank thin film. However, the upper switching field is
significantly above the coercivity of the blank film. This suggests the
presence of magnetic states with enhanced magnetic anisotropy, be it from
shape, strain, or interface coupling. The $R(H)$ curves for high-resistance
junctions are noisy and often show preferred multiple discrete resistance
values, suggesting the switching of multiple magnetic domains inside the
electrodes. Over the many junctions we have measured to date, a wide variety
of magnetic behaviors have been seen. Fig.\ref{fig10} shows some other $R(H)$
loops with complex magnetic structures. Apparently, much better understanding
and control of the electrode's magnetic state is necessary.

It is also possible that part of the $R(H)$ loop complexities comes
from magnetic defect-sites inside the barrier, rather than from the
magnetic states of the electrodes alone. Barrier-related magnetic
defects has been shown in small ($<$80nm diameter) Ni-NiO-Co junctions 
to cause magnetic field-dependent fluctuations that can
lead to large apparent magnetoresistance at low frequencies \cite{384}.

\subsection{Bias dependence of junction resistance}

Non-linear $IV$ characteristics are present in all junctions that show large
magnetoresistance. Following definitions similar to that from Fig.\ref{fig7}%
, at a given bias current for a junction we define $V_{high}$ and $V_{low}$
as the high- and low-voltage state corresponding to $R_{high}$ and $R_{low}$%
. We trace $V_{high}$ and $V_{low}$ as a function of bias current, and the
resulting two branches of the $IV$ characteristics and the bias-dependent
magnetoresistance ratio $C_R$ are shown in Fig.\ref{fig11}. The two branches
of $IV$s show pronounced non-linearity. At the same time the
magnetoresistance ratio shows a bias-dependence, with the
contrast $C_R$ decreasing as junction bias is increased. This behavior is
similar to what is observed in metal-based magnetic tunneling junctions \cite
{223}, although the characteristic voltage scale for the reduction of
magnetoresistance is at least a factor of 3 to 5 smaller than in the case of
metal junctions such as CoFe/Al$_2$O$_3$/Co \cite{388}.

The bias-dependent junction transport can be further examined by plotting
the differential junction conductance as a function of bias voltage. This is
shown in Fig.\ref{fig12}, where the specific differential junction
conductance is plotted as a function of junction bias, showing a relatively
sharp conductance minimum at low-bias riding on top of a more gradually
rising high-bias conductance slope. The origin of this characteristic is not
well understood. It is more complex than one would expect from a simple
metal-insulator-metal tunneling junction as the Simmons tunneling model \cite
{349} would predict (dotted lines). More knowledge about the barrier and the
interface states of the CMR\ materials is necessary.

\subsection{Size-dependence and inhomogeneity}

CMR\ trilayer junctions fabricated to date show large amount of
inhomogeneities. Junction conductance in most cases do not scale with
junction area in any simple way. This is evident in data presented in Fig.%
\ref{fig12}. If the junction conductances were to scale with the area, one
would expect all specific conductance curves to be roughly the same in
value. Instead in Fig.\ref{fig12} the specific conductance increases with
increasing junction size. Fig.\ref{fig13} shows two sets of junction
resistances as a function of pillar size for two rows of junctions on one chip. 
A clear cross-over from high junction resistance to
low junction resistance can been seen as the junction dimension is increased
beyond a cross-over region between 2 and 10$\mu $m. This indicates
non-uniform conduction over the junction area, where high conductance paths
are small in area compared to junction sizes, and they distribute with a
mean distance on the order of 2 to 10$\mu $m. Pinhole conduction is a
possibility, at least for junctions of the size above the cross-over
dimension.

The MR of manganate junctions appears to correlate to junction
resistance. Fig.\ref{fig14} shows data obtained on several manganate
junction chips. The junction magnetoresistance ratio $C_R\,$ is plotted
against the specific junction resistance on a log-log plot. These data
appear bunched into a cone boardered on the high MR\ end by a line of slope $%
1$, and on the low MR\ end by a line of slope of about $0.36$. A
slope-equal-to-1 corrlation could indicate the presence of a parallel shunt
that is magnetoresistively inactive. The meaning of the lower-bound line is
not clear.

The presence of inhomogeneity further complicates the nature of conduction
across these CMR\ junctions. Although the non-linear $IV$s from these
devices can be roughly described by the Simmons tunneling formula, it is
dangerous to conclude that such fits lead to any insight of the barrier
physics or the nature of the junction. Fig.\ref{fig15} shows a collection of
the apparent barrier thickness $t$ and barrier height $\phi $ extracted from
fitting Simmons model to various CMR junction $IV$s obtained at 4.2K. The
values of $t,\phi $ appears to form a correlation defined by $t\sqrt{\phi }%
=constant$. This apparent correlation as it turns out is a sure sign that we
are not looking at clean tunnling junctions but rather are just
parameterizing a complex non-linear device with a functional form of the
Simmons formula. A more quantitative analysis is given in the appendix.

\section{Summary and future challenges}

Large low-field magnetoresistance is demonstrated in CMR\ trilayer junctions
at temperatures below 130K. For the first time in CMR materials system a
large magnetoresistance of a factor of 5 is observed in an applied field
around $100$ Oe. This provides an existence proof that CMR\ effect can be
obtained in low-fields. Non-linear $IV$ characteristics and bias-dependent
magnetoresistances have been observed in CMR\ trilayer junctions. These
behaviors are qualitatively similar to what was observed in
metal-insulator-metal spin-dependent tunneling systems. Much of the physics
for CMR junctions remains to be explored. The major experimental challenge
is to master the materials science in the preparation of a well controlled,
well characterized barrier interface, so that transport over the junction
region could be more uniform, which will then allow more quantitative
examination of the transport process.

The control of the electrodes' magnetic state requires more study. Better
control of the shape of the junction pillar is needed to assure a
reproducible micromagnetic boundary condition. One particular issue is the
control of the amount of over etch during the fabrication of the pillar, as
illustrated in Fig.\ref{fig1}(b). Because the over-etch step height
determines to a large degree the amount of anti-parallel coupling the two
FM\ electrodes have. Also necessary is a controlled pre-setting of the
magnetic easy-axis for the manganates. In most devices studied so far, the
magnetic easy-axis is uncontrolled. That contributes to the observed
complexity of magnetic states. Further, exchange-biasing of the manganate
electrode could be looked into, since that could significantly simplify the
micromagnetics of the trilayer junction structure, and make its response to
applied magnetic field more predictable. The control of magnetic state is
particularly important in CMR\ junctions. Because of its large
magnetoresistance, detailed transport study is possible only when there is a
stable and well-controlled magnetic state.

The low-field magnetoresistance in trilayer CMR\ junctions tends to vanish
at temperatures around $130$ to $150$K, well below the Curie temperature of
the electrodes' which is around $350$ to $370$K. The reason for this is not
understood at present.

From an applications point of view, one would like to extend the temperature
range for large magnetoresistance. Room temperature operation of large
low-field CMR\ has not been demonstrated at this point, although CMR\
effects of $C_R\sim 60\%$ in $5$ tesla has been seen in La$_{0.67}$Ba$%
_{0.33} $MnO$_3$ epitaxial thin films \cite{71}. Materials stability, device
yield and compatibility to existing circuit fabrication processes are all
issues that need to be resolved.

\begin{acknowledgments}
We wish to thank Stuart Parkin, Jammie Kaufman at IBM\ Almaden Research
Center, John Slonczewski, Bill Gallagher, Arunava Gupta, Lia Krusin-Elbaum,
Peter Duncombe, Bob Laibowitz, Daniel Lopez, John Kirtley, Chang Tsuei,
Roger Koch, Robin Altman, Steve Brown, John Connolly at IBM\ Research 
Yorktown Heights;
Dan Lathrop, Rob Matthews and Steve Haupt from Quantum Magnetics Inc.,
Xinwei Li, G. Q. Gong, Yu Lu and Gang Xiao from Brown University for helpful
discussions and for assistance at various stages of the experiment.
\end{acknowledgments}

\begin{appendix}

\section{On fitting using Simmons tunneling model}

Simmons description \cite{r1,r1b,r1c,r2} of tunneling $IV$ with symmetrical
barrier can be written as: 
\begin{equation}
J\left( V,t,\phi \right) =\eta \left[ \left( \phi -\frac{qV}2\right)
e^{-A\left( \phi -\frac{qV}2\right) ^{1/2}}-\left( \phi +\frac{qV}2\right)
e^{-A\left( \phi +\frac{qV}2\right) ^{1/2}}\right]   \label{E1}
\end{equation}
where 
\[
\eta =\frac q{2\pi ht^2}
\]

and 
\[
A=\frac{4\pi t}h\left( 2m\right) ^{1/2}
\]
with barrier thickness $t$, barrier height $\phi $, and electron charge $q$.
Many people \cite{r3,362,223,r6,r7,238} have used Eqn.(\ref{E1}) to fit
experimental data, and extracted $t$ and $\phi $. There is a danger in the
literal interpretation of such $t$ and $\phi $, especially when
inhomogeneities are present.

To model a real, experimental junction one often has to consider both
parallel shunts and a decrease of effective tunneling area. Assume the
tunneling portion of the current is described by Eqn.(\ref{E1}), the
apparent junction $IV$ characteristic can be expressed as:

\begin{equation}
I\left( V,t,\phi \right) =\kappa J\left( V,t,\phi \right) +sV  \label{E3}
\end{equation}
where $\kappa \leq 1$ describes the reduction of effective tunneling area
because of barrier thickness non-uniformities, and $s$ describes the
parasitic conductance brought forth by shunts (assumed to be ohmic for
simplicity).

There is little difference in the appearance of Eqn.(\ref{E3}) compared to
Eqn.(\ref{E1}). In fact they are equivalent in functional form up to $O(V^5)$%
. Thus a two-parameter fit to Eqn.(\ref{E1}) will always give a reasonable
fit to the first order, but with modified fitting parameters $t,\phi $. The
problem is, these $t,\phi $ do not have the same physical meaning as they do
in the Simmons model Eqn.(\ref{E1}).

To see this let us create a situation where a real junction can be
represented by an ideal Simmons junction with $t_o,\phi _o$ defined
according to Eqn.(\ref{E1}). Assume the real junction has parasitic $\kappa $
and $s$ and its $IV$ can be described by Eqn.(\ref{E3}). What happens if we
use the ideal junction formula, Eqn.(\ref{E1}), to fit the real junction's $%
IV$ as defined by Eqn.(\ref{E3})? The fit will give an apparent $t,\phi $
described by the minimization condition of:

\begin{equation}
E_r=\int_{-V_o}^{V_o}\left[ J\left( V,t,\phi \right) -I\left( V,t_o,\phi
_o\right) \right] ^2dV  \label{E4}
\end{equation}
where $V_o$ is the voltage range of measurement.

At low-bias Eqn.(\ref{E1}) can be expanded to $O(V^5)$ to give:

\begin{equation}
J\left( V,\alpha ,\gamma \right) =\alpha V+\gamma V^3+O(V^5)  \label{E2}
\end{equation}
where

\begin{equation}
\left\{ 
\begin{array}{l}
\alpha \left( t,\phi \right) =\frac 12\eta qe^{-A\sqrt{\phi }}\left( A\sqrt{%
\phi }-2\right) \\ 
\gamma \left( t,\phi \right) =\frac 1{192}\eta q^3e^{-A\sqrt{\phi }}\frac
A{\phi ^{3/2}}\left( A^2\phi -3A\sqrt{\phi }-3\right)
\end{array}
\right.  \label{E2b}
\end{equation}

Since Eqn.(\ref{E1}) is equivalent to Eqn.(\ref{E2}) to $O(V^5)$, it is easy
to see the minimization of Eqn.(\ref{E4}) leads to:

\begin{equation}
\left\{ 
\begin{array}{l}
\gamma =\kappa \gamma _o \\ 
\alpha =\kappa \alpha _o+s
\end{array}
\right.  \label{E4b}
\end{equation}
with $\alpha _o$ and $\gamma _o$ defined by $t_o,\phi _o$ through Eqn.(\ref
{E2b}). So the real junction will also fit the ideal Simmons model Eqn.(\ref
{E1}) reasonably well, but with an off-set to parameters $\alpha ,\gamma $,
thus to $t,\phi $. It is therefore very misleading to use the definition of
Eqn.(\ref{E1}) to interpret $t,\phi $ and infer from that anything about the
barrier height and barrier thickness.

What makes matter worse is such fits based on Eqn.(\ref{E1}) seem to give
reasonable values for $t,\phi $. The reason behind this is the strongly
exponential dependence of the apparent conductance on $A\sqrt{\phi }$ in
Eqn.(\ref{E1}). Many non-linear odd-function $IV$ characteristics can be
expanded to $V^3$ and be assigned an $\alpha $ to the junction conductance $%
s $. An $s$ thus defined would give a value to $A\sqrt{\phi }$ according to
Eqn.(\ref{E2b}). Assume $\kappa =1$ for simplicity,

\begin{equation}
e^{-A\sqrt{\phi }}=\frac 2{A\sqrt{\phi }-2}\left( \frac s{\eta q}\right)
\label{E6}
\end{equation}
The value of $\eta q=\frac{q^2}{2\pi ht^2}=6.2\times 10^{14}\left( \frac \AA
t\right) ^2$ $\Omega $m$^2$ is large compared to typical conductance $s$
(usually around $10^6\Omega $m$^2$) for any reasonable barrier thickness $t$%
, therefore it is self-consistent to assume a large $A\sqrt{\phi }$ limit.
There the dependence of $A\sqrt{\phi }$ on $s$ is logarithmic, 
\begin{equation}
A\sqrt{\phi }=\ln \left( \frac{\eta q}s\right) +\ln \left( \frac{A\sqrt{\phi 
}-2}2\right) \approx \ln \left( \frac{\eta q}s\right)  \label{E7}
\end{equation}
and to the first order self-consistently gives 
\begin{equation}
t\sqrt{\phi }=\frac h{4\pi \sqrt{2m}}\left[ \ln \left( \frac 12\frac{q^2}{%
\pi hst^2}\right) +\ln \left( 2\pi \frac th\sqrt{2m\phi }-1\right) \right]
\sim \frac{\xi h}{4\pi \left( 2m\right) ^{1/2}}  \label{E8}
\end{equation}
where $\xi \,$is a numerical factor around $10\sim 20$, related to a
reasonable thickness estimate by:

\begin{equation}
\xi \approx 2\ln \left( \frac q{t\sqrt{2\pi hs}}\right)  \label{E9}
\end{equation}
Take our experiment for example. Our $s\approx 10^6$ $\Omega m^2$, a
reasonable range would be $t\sim 30$ \AA , or $\xi \sim 13$.

This can also be observed by numerically evaluating Eqn.(\ref{E8}). To do so
we define $t=t_1\times 10^{-10}$ m, $\phi =\phi _1q$ J, and using $%
h=6.626\times 10^{-34}$ J sec, $q=1.602\times 10^{-19}$ coulomb, $%
m=9.31\times 10^{-31}$ kg, $s=10^6$ $\Omega $m$^2$, we have for $t_1$ in
\AA\ and $\phi _1$ in eV:

\[
t_1\sqrt{\phi _1}=19.541-1.9309\ln t_1+\allowbreak .96547\ln \left( .51791t_1%
\sqrt{\phi _1}-1.0\right) 
\]
that gives

\begin{equation}
t_1\sqrt{\phi _1}\sim 16\mbox{\AA (eV)}^{1/2}  \label{E10}
\end{equation}
which sets one constrain. A specific set of $t,\phi $ is then obtained by
moving along this constrain (to satisfy linear conductance value) until it
finds an approximate descrition for the leading non-linear term as described
by $\gamma =\kappa \gamma _o$ in Eqn.(\ref{E4b}). Different junctions and
samples may have different ratio for first and third order expansion
coefficients, that leads to a set of apparent $t,\phi $ that scatters
roughly along the constrain Eqn.(\ref{E8}). This is what we saw from our
experimental fits, as shown in Fig.\ref{fig15}.

So be very careful when using Simmons formula to fit experimental $IV$
curves. If there's inhomogeneity or parallel shunt conductances present, the
apparent barrier height and thickness do not represent the values for the
tunneling barrier.

\end{appendix}



\newpage

\begin{figure}[tbp]
\caption{A cross-section transmission electron micrograph showing the
epitaxial registration of lattice fringes between the LSMO\ electrodes and
the STO\ barrier layer. Figure reproduced from ref.\protect\cite{191}}
\label{fig2}
\end{figure}

\begin{figure}[tbp]
\caption{A SEM\ micrograph of a manganate trilayer device. The center
junction size is $2\mu m\times 4\mu m$. The slightly jagged edges are due to
lift-off of the SiO$_2$ insulation layer. Shape of the actual junction
pillar is better defined than what can be seen.}
\label{fig4}
\end{figure}

\begin{figure}[tbp]
\epsfxsize=3.5in
\epsfbox{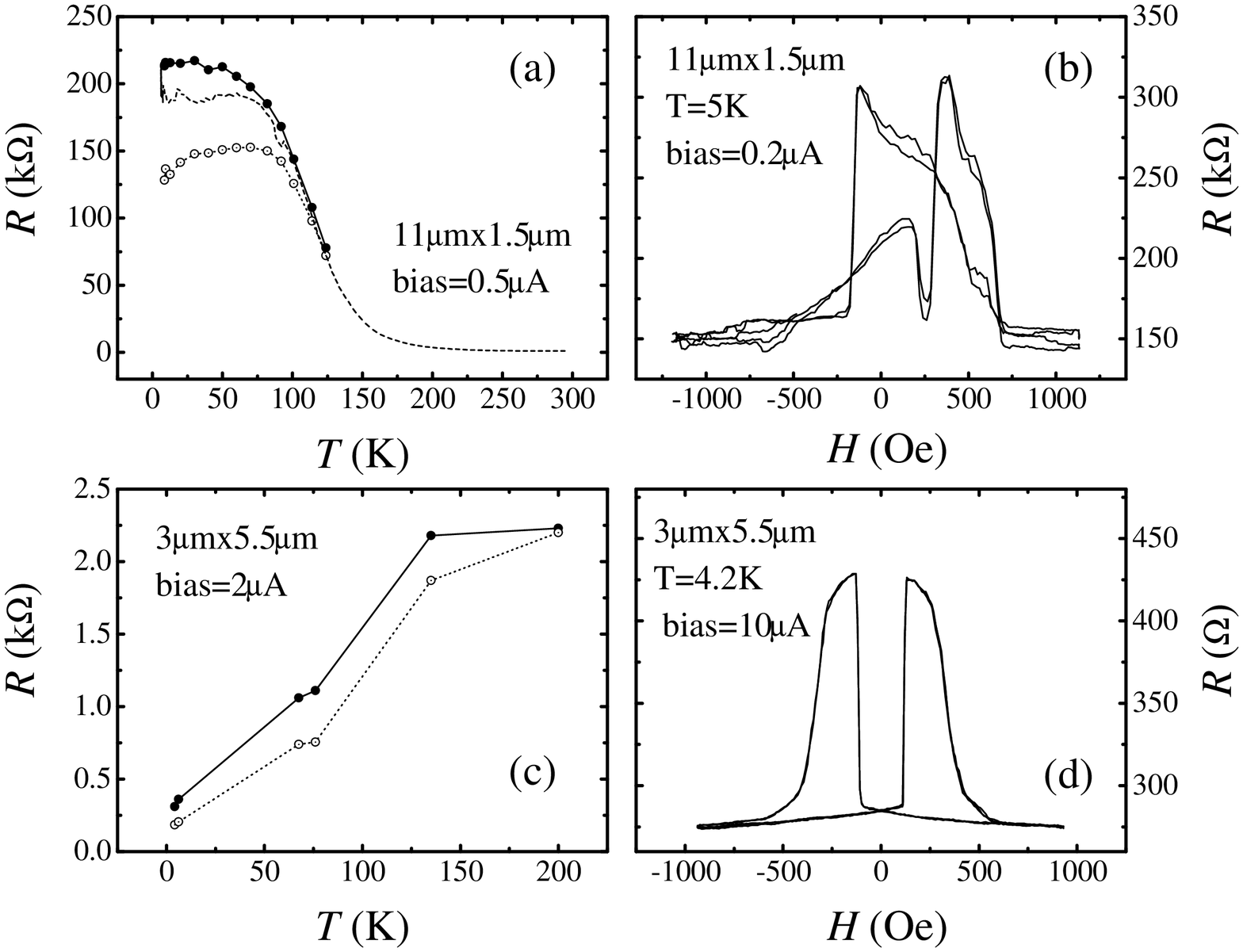}
\caption{Two types of junction behaviors under varying temperature and
magnetic field. (a) High resistance junction. Resistance increases with
decreasing temperature. Large magnetoresistance at low temperature. The 
two branches of curves show the resistive-low and -high states of the
junction. Dashed line is zero-field cooling results. (b)
Resistance vs. field loops of a high resistance junction, showing complex
switching behaviors. (c) Low-resistace junction, resistance tends to
decrease with decreasing temperature. (d) Field dependence of a
low-resistance junction. Figure reproduced from ref.\protect\cite{364}}
\label{fig5}
\end{figure}

\begin{figure}[tbp]
\epsfxsize=3in
\epsfbox{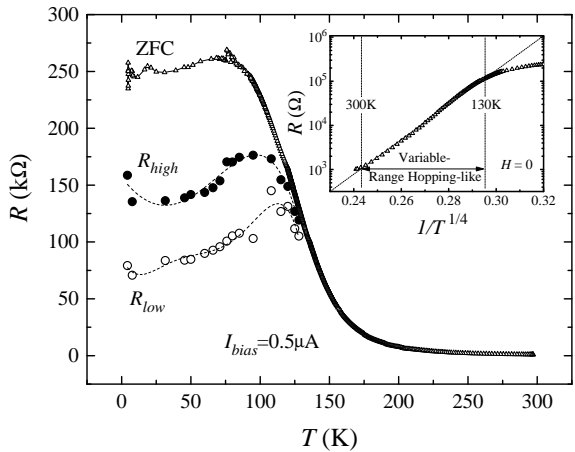}
\caption{Temperature dependence of an LSMO/STO/LSMO\ junction. We noticed
that for some junctions such as this one, the initial zero-field cooled
trace gives a higher resistance value at low temperature than subsequent
measurement cycles give. Inset: junction resistance plotted to show the $%
T^{1/4}$ scaling relation in the high temperature region between 130K and
room temperature. Figure reproduced from ref.\protect\cite{293}}
\label{fig6}
\end{figure}

\begin{figure}[tbp]
\epsfxsize=3in
\epsfbox{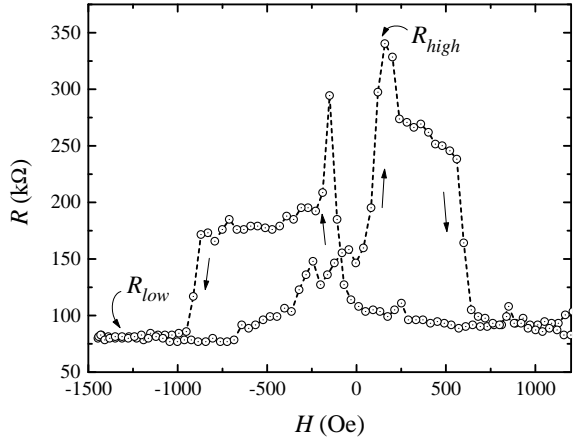}
\caption{Magnetic field dependence of junction resistance at 4.2K, showing
pronounced low-field spin-dependent behavior. A maximum resistance change by
about a factor of 5 is observed in this junction. The magnetic field in this
measurement was swept at the frequency of 1.321Hz. Figure reproduced from
ref.\protect\cite{293}}
\label{fig7}
\end{figure}

\begin{figure}[tbp]
\epsfxsize=3in
\epsfbox{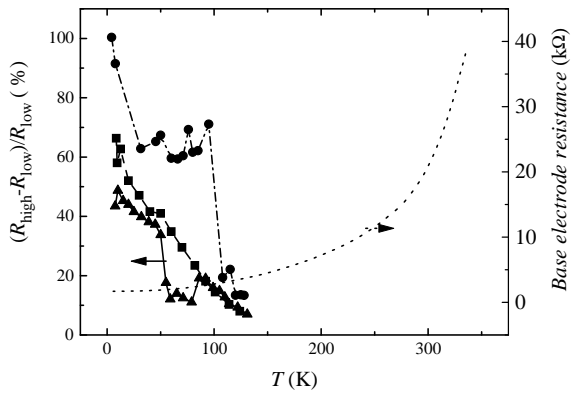}
\caption{Temperature dependence of magnetoresistance of some CPP\ junctions.
Right: temperature dependence of the base-electrode's resistivity. Figure
reproduced from ref.\protect\cite{364}}
\label{fig8}
\end{figure}

\begin{figure}[tbp]
\epsfxsize=3in
\epsfbox{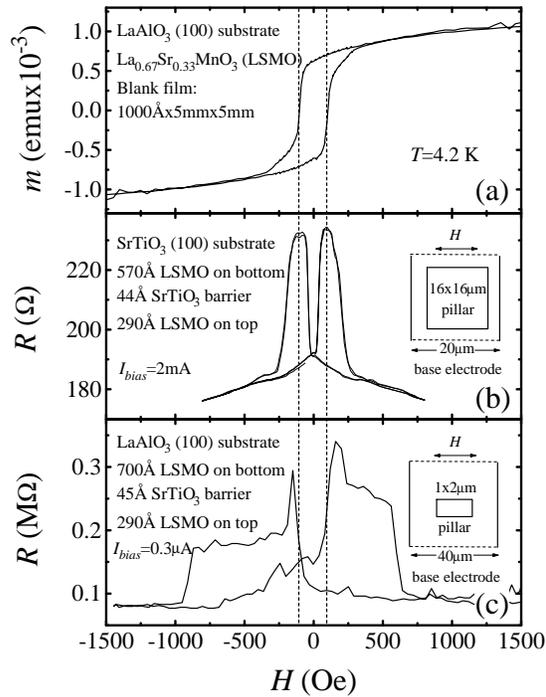}
\caption{Comparing the $R(H)$ curves of devices to magnetic hysteresis from
a blank film. (a): Magnetic hysteresis loop of a blank film. (b): $R(H)$\
loop of a low-resistance junction showing similar switching field as blank
film's $H_c$. (c) High resistance junction. The lower switching field
corresponds well to blank film's $H_c$, while the upper switching field is
well above $H_c$, indicating additional magnetic interaction is present for
magnetic states within the pillar. The insets in (b) and (c) show the
geometry of the electrodes for the particular junctions and the relative
field orientation in each case. Figure reproduced from ref.\protect\cite{364}
}
\label{fig9}
\end{figure}

\begin{figure}[tbp]
\epsfxsize=3in
\epsfbox{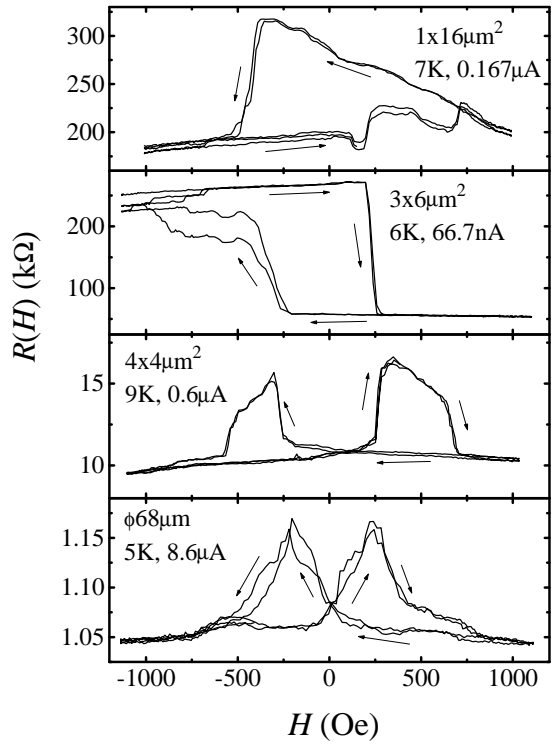}
\caption{Some more examples of field-dependent resistance of various
junctions. Junction size and measurement parameters are shown in the panels.
Figure reproduced from ref.\protect\cite{364}}
\label{fig10}
\end{figure}

\begin{figure}[tbp]
\epsfysize=3in
\epsfbox{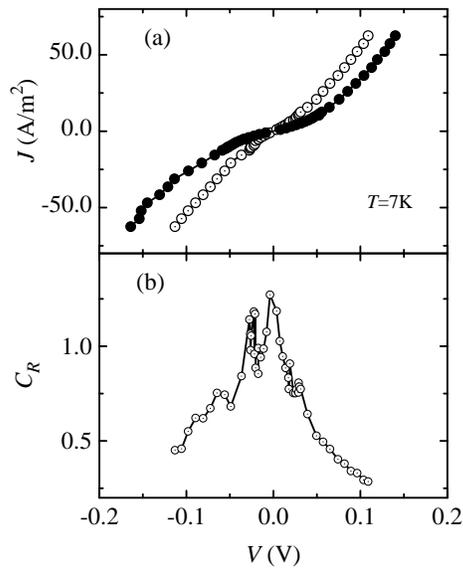}
\caption{(a) Two branches of the $IV$ characteristics of a junction showing
large magnetoresistance, corresponding to the junction's $R_{high}$ and $%
R_{low}$ state, respectively. Junction size is $11\mu $m$\times 1.5\mu $m.
(b) The magnetoresistance ratio, $C_R$, as a function of junction bias
voltage, showing a decrease of MR\ as junction bias is increased.}
\label{fig11}
\end{figure}

\begin{figure}[tbp]
\epsfysize=3in
\epsfbox{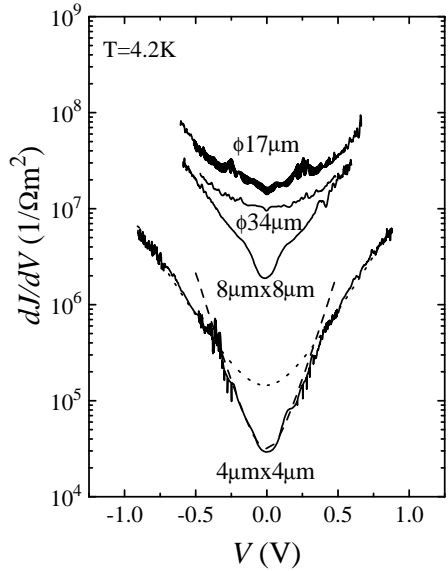}
\caption{Bias-dependent differential conductance of LSMO/STO/LSMO junctions.
The nominal barrier thickness is $39$\AA . The normalized conductance
decreases with decreasing junction area, suggesting inhomogeneity over a
length scale of several $\mu m$. The lower conductance junctions are
generally the ones that give large magnetoresistance. Notice the pronounced
low-bias conductance dip. Dashed curves are Simmons model fits to the bottom
curve. The low-bias fit gives $t=30.6$\AA\ and $\phi =0.703$eV; the
high-bias one gives $t=21.6$\AA\ and $\phi =1.32$eV. These parameters should
not be interpreted as a reflection of the barrier physics. The reason for
this is discussed further in the later part of this section. Figure
reproduced from ref.\protect\cite{364}}
\label{fig12}
\end{figure}

\begin{figure}[tbp]
\epsfxsize=3in
\epsfbox{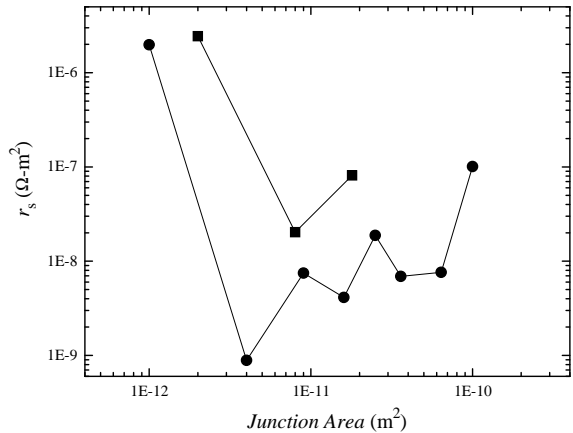}
\caption{Specific junction conductance as a function of junction area for 
different sets of junctions on the same chip. A
clear break from constant conductance is seen for devices smaller than $\sim 
$5 $\mu $m, indicating the presence of inhomogeneities over several microns.
Similar behavior has been seen on all chips studied. }
\label{fig13}
\end{figure}

\begin{figure}[tbp]
\epsfxsize=3in
\epsfbox{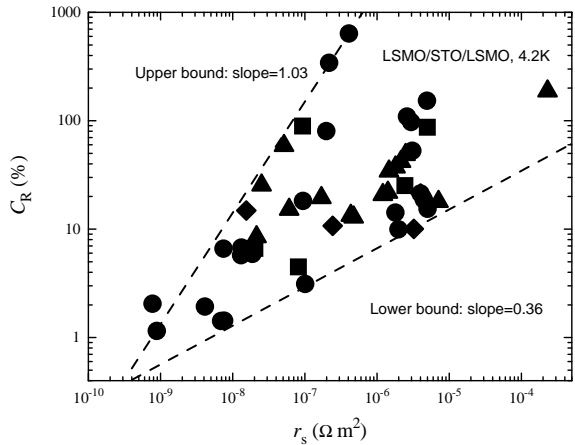}
\caption{A summary plot of manganate junctions' MR as a function of specific
junction resistance $r_s$. Data were gathered from junctions on same chip as
well as from different chips with varying deposition conditions and
different nominal barrier thicknesses. The data fall within a cone of upper
slope $1$ and lower slope around $0.36$. A slope $1$ scaling on such plot
might indicate the presence of a distribution of parallel shunt conductances
that are magnetoresistively inactive, which might explain the upper bound.
The slope of the lower bound is not understood.}
\label{fig14}
\end{figure}

\begin{figure}[tbp]
\epsfxsize=3in
\epsfbox{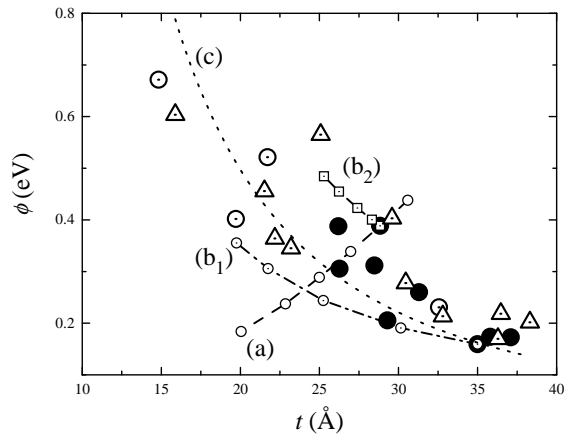}
\caption{Fitting to Simmons formula Eqn.(\protect\ref{E1}) of many CMR junctions'
4.2K $IV$ gave this distribution of the variables $t,\phi $. The apparent
correlation between $t$ and $\phi$ is most likely due to the presence of
parallel shunt conductance (if the principal non-linear conductance does
come from tunneling). A distribution of effective area $\kappa$ between $1.0$
and $10^{-4}$ leads to a set of varying $t,\phi$ as shown in curve (a),
which starts with $t=28.84$ \AA, and $\phi = 0.3885 $ eV at $\kappa = 1 $. A
distribution of parallel shunt conductance by a factor of $10$ causes $%
t,\phi $ to spread as illustrated in curves (b$_1$) and (b$_2$). Curve (c)
is an example of a constant product of $t\phi ^{1/2}=$ 14.1 \AA $\mbox{(eV)}%
^{1/2}$.}
\label{fig15}
\end{figure}

\end{document}